\documentclass[12pt]{article}
\usepackage[english]{babel}
\usepackage[pdftex]{graphicx}
\usepackage{hyperref}
\hypersetup{
    colorlinks = true,
    linkcolor = blue,
    anchorcolor = blue,
    citecolor = blue,
    filecolor = blue,
    urlcolor = blue,
	pdfauthor={some author},
	pdftitle={eye-catching title}
    }
\usepackage{amssymb,amsmath}
\usepackage{url}
\begin{document}

\title{The reification of the word against science\footnote{This is the English translation and revision of an essay published in Italian: L. Foschini, ``La reificazione della parola contro la scienza''. \emph{La Citt\`a del Secondo Rinascimento}, n. 99, Giugno 2022, p. 13--15.}}
\author{Luigi Foschini\footnote{Brera Astronomical Observatory, National Institute of Astrophsyics (INAF), 23807 Merate, Italy. Email: \texttt{luigi.foschini@inaf.it}.}}
\date{\today}
\maketitle

\begin{abstract}
Research in foundations of physics is stagnant, as claimed by many scientists during the last years. I suggest that one reason might be the reification of language, particularly of mathematics, which favour the search for answers to wrong questions.  
\end{abstract}

Many scientists assert that research in theoretical and/or fundamental physics is declining or stagnating. Jacome Armas \cite{ARMAS} interviewed 37 renowned scientists about quantum gravity and asked them what they think to be the biggest breakthrough in theoretical physics in the last 30 years\footnote{The interviews were collected between 2011 and 2020, so the last 30 years refer to the period starting from eighties.}. Twelve scientists (37.5\%)\footnote{Armas did not pose this question to five scientists.} replied that no breakthrough was done. This is a significant part. In addition, among the scientists interviewed by Armas who replied indicating some breakthrough, four referred to experiments or observations useful for the theory, rather than theory itself. Surely, during the latest three-four decades, there were important experimental or observational results, such as the $W^{\pm}$, $Z$, and Higgs bosons, the gravitational waves, the quantum teleportation, the acceleration of the expansion of the Universe, but the theory was developed many years earlier in all the cases. Therefore, if one adds also these four answers to the twelve cited above, the number of scientists reporting no breakthrough in theoretical physics increases to 50\%. The remaining 50\% voted for the string theory, with particular reference to AdS/CFT duality. 

Sabine Hossenfelder \cite{HOSSENFELDER18B,HOSSENFELDER19} wrote about this stagnation many times and the title of her book is exemplary: lost in math \cite{HOSSENFELDER18A}. Hossenfelder denounces, among other things, the dramatic lack of philosophy among her theoretical colleagues, which leads them to consider exclusively the mathematical technique and its beauty. String theory is one clear example, but it is also sufficient to take a look at the works published in recent times to note a complete detachment from physical reality, not to mention the closure in a world of fantasy. Almost all the essays deal with questions such as: what are the consequences if this parameter has the value $x$ instead of the actually measured value $y$? The so-called multiverse is the most striking example: failing to explain with technique alone (obviously) why we live in the present Universe, the existence of infinite universes has been postulated, each with different values of the fundamental constants (e.g. \cite{TEGMARK}). For example, there will be somewhere a universe where the electric charge of the electron has a value different than the measured one, greater or smaller, significantly different or not. Therefore, the Universe would be as it is simply because we live in one of the infinite universes available, the one in which the fundamental constants have the values we have measured. Another way to remain stuck into fantasy is to invent new absurd useless particles, and to neglect decades of failures to detect them \cite{HOSSENFELDER22}.

Anyone who thinks that these are issues light years away from everyday life, must understand that this stagnation does not only concern the fundamental physics, but also the technology that derives from it. The science journalist Michael Hanlon \cite{HANLON} wrote that contemporary technology is nothing more than a refinement of what has been developed in the postwar period, in that period called the Golden Quarter, which runs approximately from 1945 to 1971. For example, smartphones are based on the updating and assembly of technologies invented in that period: the transistor (1947), the integrated circuit (1958), the touch screen (1965), the lithium batteries (1970), the microprocessor (1971)\footnote{See Federico Faggin's personal recollections \cite{FAGGIN}}. In addition, these technologies are all based on quantum mechanics developed in the first half of the 20th century. There is a lot of talk about investing in scientific research, but -- \emph{de facto} -- people invest in refining old technologies. Reality is that theoretical, curiosity-driven research is considered marginal, if not useless, because it is not immediately productive. 

Giorgio Israel \cite{ISRAEL} wrote that this is the result of the spread of scientism, which is based on post-modern, anti-humanist and materialistic thought. Science is a technique plus a philosophy: the science-zealot\footnote{Italian language has two different words to distinguish a scientist (\emph{scienziato}) from a fanatic (\emph{scientista}). English language has not this divide, although the word \emph{scientism} indicates an excessive trust in science. Therefore, I adopted the composed word \emph{science-zealot} to indicate a fanatic scientist.} is not only convinced to do science by renouncing philosophy, but also thinks that technoscience must be applied to all the human knowledge. To understand what monstrosity this mutilated science can generate, it is sufficient to remind the scientism-based programs of palingenesis of the human species operated by the Nazi-Fascist and Communist dictatorships in the twentieth century. Nonetheless, scientism still continues to rage today, even if, like a virus, has changed its form, re-proposing itself in a sort of laic religion, such as Rousseau's naturalism. The trouble is that scientific institutions continue to close their eyes, while -- as Israel notes -- there is a desperate need to admit that many problems in science are caused more by internal than external enemies. An underestimated consequence of the diffusion of scientism is the destruction of people's trust in science, as we saw during the recent pandemic with the spread of various quackery alternatives to medicine.

In theoretical/foundational physics, the problems generated by technoscience manifest themselves mainly with the reification of mathematics. Mathematics was for Galilei the language used by God to write the book of nature, while for Bohr there was not only the mathematical language, but any language. Physics is about what we can \emph{say} about nature, not what nature \emph{is}, as the Danish physicist wrote (cited in \cite{PETERSEN}). On the opposite, neo-platonists like Max Tegmark \cite{TEGMARK} claimed that the Universe is mathematical\footnote{Incidentally, this is a misunderstanding that even Israel runs into, when he attributes to Galilei the view that the world is mathematical.}. There is a huge difference between what Galilei wrote, when he spoke about mathematics as a language, and today's vision of a mathematical world, which is the reification of the mathematical language. It would be a bit like saying that the world is Italian or English or any other language or dialect: that's simply ridiculous!

\begin{figure}[!t]
\begin{center}
\includegraphics[scale=0.3]{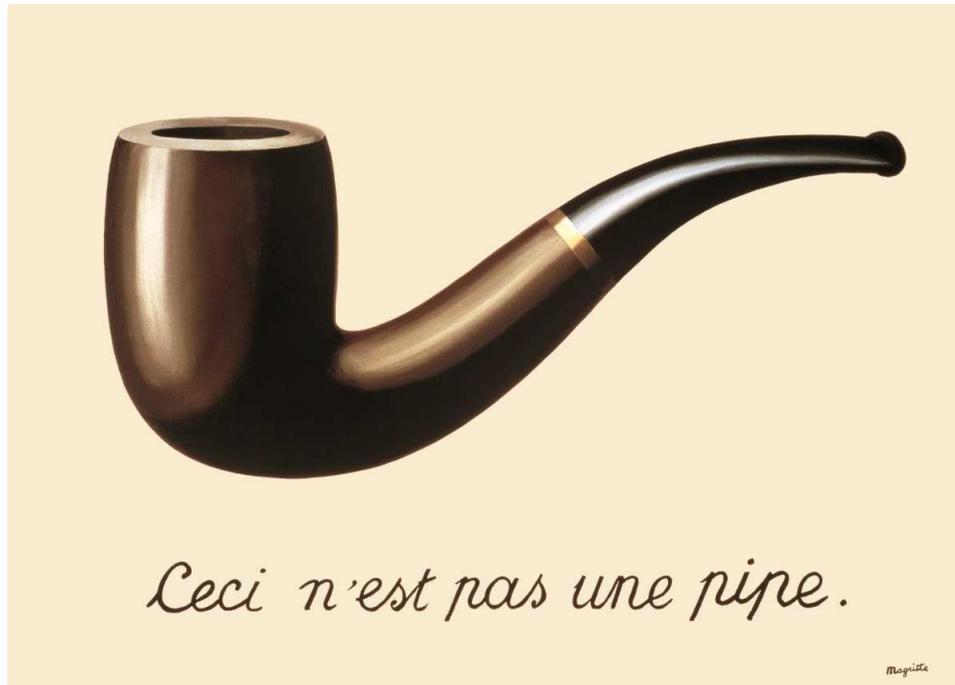}
\caption{Ren\'e Magritte, \emph{La Trahison des images}. Oil on canvas, (1929).}
\label{magritte}
\end{center}
\end{figure}

The Italian philosopher Giovanni Vailati \cite{VAILATI} wrote that the starting point of science is the language, mathematics or any other set of signs, not the things. We associate words with things, we make assumptions and simplifications, we try to better specify the linguistic conventions to represent physical reality and to build the tracks on which to make logical inferences, because all we can do is to build an isomorphism between the reality and the logic of the adopted language. Thinking that science is reality, that the world is mathematical, it means to reify the mathematical word or, as Ludwig Wittgenstein wrote, transforming a substantive into a substance \cite{WITTGENSTEIN}. Think about the famous painting \emph{La Trahison des images} (1929) by Ren\'e Magritte (Fig.~\ref{magritte}), in which the painter drew a pipe and wrote: ``Ceci n'est pas une pipe''. This should be quite evident: the drawing is not the thing, like words are not the things they refer to. On the contrary, the reification of mathematics is like to be convinced that the painting \emph{is} a pipe! This is what the reification of mathematics means. Just to cite a recent example, in an editorial note published on \emph{Nature Physics} is written: ``However, the dynamics of a wavefunction collapse has never been observed and involves an uncomfortable division between the classical observers and the quantum systems they are measuring'' \cite{NATURE}. And it will never be observed, because, as Asher Peres wrote, ``quantum phenomena do not occur in a Hilbert space, they occur in a laboratory'' \cite{PERES}. The wavefunction is just a mathematical symbol in a relationship with something physical, it is not a physical object.  Sabine Hossenfelder wrote that ``the reason for the current lack of progress may be that we focus on the wrong questions'' \cite{HOSSENFELDER18A}. Exactly, and the reification of mathematics is the freeway for wrong questions: people are lost in searching for pipe paintings, instead of looking for real pipes. Are we then surprised that physics is stagnating?

Again Vailati wrote that our idea of real existence is very ephemeral \cite{VAILATI}: after all, by saying that a thing really exists, we believe that if another human being interacted with it, he would have the same sensations as us (which is not obvious: think, for example, at a color blind person). Since it is not possible to feel what another human being feels, we have the language that allows us to interact with each other. With macroscopic objects, the interaction is relatively simple: even with all the ambiguities of languages, a table is easily identifiable and recognizable. But when dealing with abstract concepts, such as love or freedom, problems already arise. For this reason, in science, there is often discussion about definitions and an abstract language such as mathematics has been built.

As already written, what we can aspire to is to build an isomorphism with reality. Again, as long as there are macroscopic bodies, as in classical physics, everything is relatively fine. Problems re-emerge with experiences beyond human sizes, such as the very small (quantum mechanics) or the very fast (relativity). However, even the success of classical physics, which generated determinism, positivism and therefore scientism, in reality, hides pitfalls. Laplace (1796) said that if one knows the position and impulse of all the bodies in the Universe, then it would be possible to predict the future. Even today, from time to time, we read statements by scholars who affirm that this would be possible, although only in theory. This would be true only with the reification of mathematics, if the mathematical word were the physical body. However, what we indicate with the quantities position and impulse are only two of the infinite properties of physical bodies, and we consider these two because they are useful for a certain study (mechanics). In other words, we have neglected all the other properties, because they are not essential to our purpose, which was to calculate the dynamics of a macroscopic physical body\footnote{See also the importance of statistical regularities in supporting some kind of predictability \cite{MYRVOLD}.}. We also neglect time: to measure the above cited quantities, it is necessary to perform two experiments, one after the other, but we assume that the outputs can be considered as done at the same time. This is no more possible in quantum mechanics, because the perturbation induced by the measurement are no more negligible. Also macroscopic bodies are not free from these issues: chaos theory has shown that, in dynamics over very long times, even what we have neglected because considered insignificant can have a significant influence (for example, the Solar System might not be stable on long time scales, see the review by Laskar \cite{LASKAR}). Therefore, determinism is not possible even in theory, simply because the word is not the thing. Ceci n'est pas une pipe! And if this is not enough, it is sufficient to see the stagnation in science and technology produced in the last half century by this ideology of science, which denies the word.

The development of science has gone hand in hand with the dematerialization of the word. I have already written a book on this topic \cite{FOSCHINI15B}\footnote{The book is in Italian, but some preliminary essays in English are available in \cite{FOSCHINI11,FOSCHINI13,FOSCHINI15A}.}, but here it is useful to recall some examples. Read the books written by Arnold Sommerfeld, one of the most important contributors to atomic physics, as well as supervisor of eight Nobel laureates and of many other outstanding physicists: you will find linguistic annotations and a great care of words. Paul Dirac, known for the equation that bears his name and anticipated the discovery of antimatter, wrote a textbook on quantum mechanics with an almost obsessive attention to words, to the point that he simply read his text in the classroom. When asked by a student to explain a concept in other words, he replied that no, it was not possible, because the selected words were the best. When the Italian mathematician Gregorio Ricci Curbastro invented the tensor calculus, he was opposed by many colleagues, who thought it was only a linguistic rehash of known ideas. However, Einstein had already tried all known ideas for gravitation when he desperately turned to Marcel Grossman, pleading for help. And Grossman suggested the so-called rehash of Ricci Curbastro, which turned out to be extremely effective for being a mere linguistic question. Think about the great debates between Einstein and Bohr, which represent the most beautiful moments of the golden age of theoretical physics: they are all linguistic discussions, \emph{gedankenexperiment}, without setting foot in the laboratory. Still, today's electronic technology wouldn't exist without that exchange of words. Read the biographies of the great scientists: more or less everyone recognizes the importance of words and how language structures the way of thinking and therefore inventiveness. Particularly Einstein, having to prepare an essay about his life, wrote: ``For the essential in the being of a man of my type lies precisely in what he thinks and how he thinks, not in what he does or suffers'' \cite{EINSTEIN}.

\section*{Acknowledgements}
I would like to thank Manfred Bucher for pointing out an error I made about Arnold Sommerfeld (he was not a Nobel laureate).

\end{document}